\documentstyle[12pt]{article}
\begin{document}
\begin{titlepage}
\title{
{\bf Interaction Hierarchy \\
}
}
{\bf
\author{
 G.K. Savvidy\\
 Physics Department, University of Crete \\
 71409 Iraklion, Crete, Greece \\
 and\\
 Yerevan Physics Institute, 375036 Yerevan, Armenia \vspace{1cm}\\
 K.G.Savvidy\\
 Princeton University, Department of Physics\\
 P.O.Box 708, Princeton, New Jersey 08544
}
}
\date{}
\maketitle
\begin{abstract}
\noindent

We analyse a new class of statistical systems, which
simulate different systems of random surfaces  on a
lattice. Geometrical hierarchy of the energy functionals
on which these theories are based produces corresponding
hierarchy of the surface dynamics and of the phase
transitions. We specially consider 3D gonihedric system
and have found that it is equivalent to the propagation
of almost free 2D Ising fermions. We construct
dual statistical system with new matchbox spin variable
$G_{\xi}$, high temperature expansion of which equally
well describe these surfaces.

\end{abstract}
\thispagestyle{empty}
\end{titlepage}
\pagestyle{empty}

{\Large{\bf 1.}}
The correspondence between spin configurations and the
surfaces of interface allows to define different theories of
random surfaces on a lattice. Ising
ferromagnet is equivalent to a system of random walks with the
energy which is proportional to the $\it length$ of the path on the
two-dimensional lattice and to the $\it area$
of the surface in three-dimensions
\cite{waerden,kramers,ising,onsager,kac,green,schultz,wegner,fanwu,kadanoff}.

 In the recent articles  \cite{wegner1,savvidy}
the authors formulated a new class of statistical systems,
whose interface energy is associated with the edges of
the interface. Spin systems introduced in
\cite{wegner1,savvidy} represent
a system of random walks, but  with
the energy which is proportional
to a total $\it curvature$ of
the path on the two-dimensional lattice
and to the $\it size$ of the surface in three dimensions
\cite{ambartzumian,savvidy1,savvidy2}. One motivation for
the study of those  statistical systems is to have
well defined physical system of random surfaces which
is relevant for the description of the gauge field
dynamics and of the QCD \cite{ambartzumian,savvidy1,savvidy2}.

In order to study the dynamics of the surfaces with
linear-gonihedric action \cite{ambartzumian,savvidy1,savvidy2}
we analyse statistical
systems with topological and area action and
compare  statistical properties of these systems
from the viewpoint of surface dynamics.
The energy functionals on which these theories are based
represent a realization of the  hierarchy of geometrical
functionals found by Steiner and Minkowski
\cite{steiner,minkowski,blaschke,chern,santalo,hadwiger,ambartzumian1,weil}.
The imposed structure has an important influence
on the physical properties of the system and produces
corresponding hierarchy of the surface dynamics and
of the phase transitions.  We have found that
3D gonihedric system is equivalent to the propagation
of string with almost free 2D Ising fermions. We construct
dual statistical system with new matchbox spin variable
$G_{\xi}$, high temperature expansion of which equally
well describe these surfaces.

{\Large{\bf2.}} The origin of the interaction hierarchy
lies in  Steiner's geometrical idea about parallel
manifold $M_{\rho}$
\cite{steiner,ambartzumian,blaschke,bla}.
His formula represents the area of the parallel surface
$M_{\rho}$, situated on the distance $\rho$ from
$M$, in terms of the area $S$, mean integral
width $A$ and the topology $T$ of
the initial convex surface $M$

$$S_{\rho} = \int_{M}R_{1}R_{2} d\omega
+ \rho \int_{M}(R_{1} + R_{2}) d\omega +
\rho^{2} \int_{M} d\omega  ,\eqno(1)$$
where $R_{1,2}$ are the principal curvatures of the
surface $M$ and $d\omega$ is the area
element of the spherical map. All
this functionals, $S(M)$, $A(M)$ and $T(M)$
properly defined for nonconvex case
\cite{ambartzumian,savvidy1,savvidy2,wegner1}
produce the systems of interacting surfaces
with a very deep hierarchical structure.

At the first level one can define topological-gonimetric
systems on the lattice, these are
1D Ising ferromagnet, 2D gonimetric
walks and 3D gonimetric surfaces
\cite{wegner1,savvidy}

$$H^{1D}_{Ising} = -\sum_{links} \sigma \sigma ;
H^{2D}_{gonimetric} = -\sum_{plaquettes}
 \sigma \sigma \sigma \sigma ;
H^{3D}_{gonimetric} = -\sum_{boxes} \sigma \sigma \sigma \sigma
\sigma \sigma \sigma , \eqno(2)$$
where the spin variables are attached to verteces of the
lattice. Using high
temperature expansion one can see that all topological
systems (2) have the same partition function and are in
a completely disordered phase \cite{savvidy}

$$-\beta f_{gonimetric} = ln(\omega + \omega^{-1}), \eqno(3)$$
where $\omega = e^{-\beta}$.

Increasing the dimension of the lattices by one and leaving
the Hamiltonians (2) without changes, one can see that the
system (2) "moves" to next level of the hierarchy and
describes now the system with gonihedric action
$A(M) = \sum_{<i,j>}  \vert X_{i} - X_{j} \vert
\cdot \vert \pi - \alpha_{i,j} \vert$ which is
proportional to the linear size of the system
\cite{ambartzumian,savvidy1,savvidy2}.

There is mutual connection between gonimetric and
gonihedric systems. Indeed in the recent article
\cite{schneider} the authors have found that the
action $A(M)$ has an equivalent representation
in terms of total curvature
$k(E)$ of the polygons which appear in the intersection of the
two-dimensional plane $E$ with the given surface $M$ .
This curvature $k(E)$ should then be
integrated over all planes $E$ intersecting the surface $M$.
This result directly connects
topological-gonimetric system in two
dimensions with gonihedric system in three dimensions.
This connection is universal for any dimensions.

Indeed the intersection of the two-dimensional
plane $E$ with the polyhedral
surface $M$ is the union of the polygons

$$P_{1}(E) , ...,P_{k}(E) \eqno(4)$$
and as it was shown in \cite{schneider}, the absolute total
curvature $k(E)$ of the  polygons (4) in the intersection is

$$ k(E) = \sum^{k}_{i=1} k(P_{i}) =
\sum_{<i,j>} \vert \pi-\alpha^{E}_{ij} \vert , \eqno(5)$$
where $\alpha^{E}_{ij}$ are the angles of the polygons (4) and they are
defined as the angles in the intersection of the two-dimensional
plane $E$ with the edge $<X_{i},X_{j}>$, that is with
the dihedral angle $\alpha_{ij} $\cite{schneider}.
Particularly for the plane
$E$ which is perpendicular to the edge $<X_{i},X_{j}>$
we will have that $\alpha^{E}_{i,j}=\alpha_{i,j}$.
The meaning of (5) is
that it measures the total revolution of the
tangent vectors to polygons
(4) $P_{1}(E) , ...,P_{k}(E)$ \cite{fenchel,borsuk,milnor,chern1}.
The summation in (5) can be extended
to all edges of the surface $M$ if
we take $\alpha^{E}_{ij} = \pi $
for the edges which are not intersected by the given plane $E$.
Finally to  get an action $A(M)$ one should integrate
the total curvature $k(E)$ (5) over all
intersecting planes $E$ \cite{schneider}

$$ A(M) = \frac{1}{2\pi}\int_{\{ E \} }k(E)dE, \eqno(6)$$
where the integral is extended over all planes$ \{ E \}$
intersecting the surface $M$. The measure $dE$ in (6) is
equal to
$$ dE = dp d\Omega \eqno(7) $$
where we define the plane $E$ by its normal vector
$\vec n = (\sin\theta\cos\phi,\sin\theta\sin\phi,\cos\phi)$
and its distance to fixed origin. $d\Omega$ denotes the
area element of the unit sphere corresponding to the end point
of the normal vector $\vec n $. In the next section we will
apply (6) to lattice surfaces.

{\Large{\bf3.}}
Singular surfaces $\{ M\}$ of interface \cite {waerden}
on a cubic lattice $Z^{3}$
can be considered as a collection of plaquettes
with the restriction that only an even number of plaquettes
can intersect at a given edge ( $2r = 0,2,4.$) and that only
one plaquette is on a given place \cite{wegner,wegner1,savvidy}.

As a first step we should find the same
representation (6) for the action
$A(M)$ on a cubic lattice $Z^{3}$. This can be easily done if we
introduce a set of planes $\{ E_{x} \}, \{ E_{y} \}, \{ E_{z} \} $
on the dual lattice $Z^{3}_{d}$ which are perpendicular to $x, y$ and
$z$ axis correspondingly.

These planes intersect a given surface $M$ and
on each of these planes we will have an image of the  surface
$M$. Every such image is represented as a collection of
closed polygons $\{ P(E)\}$ appearing in the intersection of
the plane with surface.

The energy of the surface $M$ is equal now to a
sum of the total curvatures $k(E)$ of all these polygons
on  different planes

$$ A(M) = \sum_{over~all~dual~planes \{ E_{x},E_{y},E_{z}\} }
k( E ). \eqno(8)$$
On the lattice the total curvature  $k(E)$ (5) is simply
the total number of polygons right angles. In general,
the surface on the lattice may have self-intersections and
one should associate the energy with self-intersection edges of
the surface and therefore with the self-intersections of polygons.
Depending on how we will count the right angles in the
self-intersection vertices  we
will get different theories \cite{savvidy2,wegner1,savvidy}.

In the present article we will consider the case
when the self-intersection coupling constant $\kappa$
is zero \cite{savvidy}. This means that when we compute the
total curvature $k(E)$ of
the polygons in  (8) we should ignore the right angles at
the self-intersection points.

In terms of Ising spin variables $\sigma_{\vec r}$ the Hamiltonian
of this system has the form \cite{wegner1,savvidy}

$$H^{3D}_{gonihedric} = - \sum_{\vec r,\vec \alpha,\vec \beta}
\sigma_{\vec r} \sigma_{\vec r+\vec \alpha}
\sigma_{\vec r+\vec \alpha+\vec \beta}
\sigma_{\vec r+\vec \beta}, \eqno(2a)$$
where $\vec r$ is a three-dimensional vector whose components
are integer and $\vec \alpha$ ,$\vec \beta$
are unit vectors parallel to axis.
The gonihedric system (2a) has an extra
symmetry: one can independently flip spins on any
combination of planes (spin layers).

{\Large{\bf4.}}With (8) the partition function of the
system can be written in the form

$$  Z(\beta) = \sum_{over~all~\{M\}}
exp\{-\beta \sum_{\{E\}} k(E)\}.  \eqno(9) $$
Let us represent the sum in the exponent as a product

$$  exp\{-\beta \sum_{\{E\}} k(E)\} =
\prod_{\{E\}}  e^{-\beta k(E)} =
\prod_{\{E_{z}\}}  e^{-\beta k(E_{z})}
\prod_{\{E_{y}\}}  e^{-\beta k(E_{y})}\prod_{\{E_{x}\}}
e^{-\beta k(E_{x})} \eqno(10)$$
It is true that these products are not independent for a given
surface $M$
and our goal is to express the initial energy (8) and the
product (10) in terms of
independent quantities, that is only through
the product over all two-dimensional planes in one fixed direction,
let's say through $\{E_{z}\}$.

The question is: what kind
of information do we need to know on these planes  $\{E_{z}\}$
to recover the values of the total curvature $k(E_{x})$
and $k(E_{y})$ on the planes  $\{ E_{y}\}$ and $\{E_{x}\}$
for the given surface $M$ ?

With this aim let us consider the sequence of two planes
$E_{z}^{i}$ and  $E_{z}^{i+1}$ and denote by $P_{i}$
the polygon-image of the surface $M$ on the plane  $E_{z}^{i}$
and by $P_{i+1}$ the polygon-image of $M$ on the plane
$E_{z}^{i+1}$.

With this one can compute the contribution to the total curvature
$ k(E_{x})$ which comes from the pieces of the polygon-images on the
planes $\{ E_{x} \}$ which lie  between two planes  $E_{z}^{i}$ and
$E_{z}^{i+1}$. This contribution is equal to the number of
bonds of the polygons $P_{i}$ and $P_{i+1}$ which are parallel
to the $x$ axis without common bonds.

In the same way the contribution to the $k(E_{y})$ from the polygons
on the planes $\{ E_{y}\}$ which lie between $E_{z}^{i}$ and
$E_{z}^{i+1}$ is equal to the length of the non-common bonds of
the $P_{i}$ and  $P_{i+1}$ which are  parallel the $y$  axis.

Therefore the total contribution to the curvature $k(E_{x})+k(E_{y})$
of the polygons which are on the perpendicular planes between
$E^{i}_{z}$ and $E^{i+1}_{z}$  is equal to the length of the polygons
$P_{i}$ and $P_{i+1}$ without length of the common bonds

$$ l(P_{i})+l(P_{i+1})-2 \cdot l(P_{i} \cap P_{i+1} ). \eqno(11)$$
This formula represents an important fact that the
$\it curvature$ of the polygons which lie on the perpendicular planes
$\{ E_{y}\}$ and $\{E_{x}\}$ is equal
to the $\it length$ of the polygons on the $\{ E_{z} \}$ planes

$$ \left( \begin{array}{c}
   Contribution~ to~k(E_{x})+k(E_{y}) \\
   from~planes~\{ E_{x}\},~\{ E_{y}\} \\
   between~E_{z}^{i} ~and~ E^{i+1}_{z}
   \end{array}  \right) =
   l(P_{i})+l(P_{i+1})-2 \cdot l(P_{i} \cap P_{i+1} ) .\eqno(12)$$
Where $ l(1) + l(2)-2\cdot l(1 \cap 2) = l(1\cup 2\setminus 1 \cap 2) $.
Now the energy functional (8) is reduced to an independent sum
over the polygon loops only on the $E_{z}$ planes

$$A(M) = \sum_{\{ E\}} k(E) = \sum_{\{ E_{z} \}}
k(P_{i}) + l(P_{i}) +  l(P_{i+1})
- 2 \cdot l(P_{i} \cap P_{i+1}) \eqno(8a)$$
which we will consider  as a sum of the free action
proportional to the total $\it{curvature}$ plus the $\it{lenght}$

$$A_{0}(P) = k(P)/2 + l(P) \eqno(8b)$$
and of the interaction term which is proportional to the
$\it length~of~overlapping$ of the polygon loops $P_{i}$
and $P_{i+1}$

$$A_{int}(P_{i},P_{i+1}) = - 2\cdot l(P_{i} \cap P_{i+1}). \eqno(8c) $$
With this formula one can represent
the product (10) in the form

$$ \prod_{\{ E \} } e^{-\beta k(E)}
=   \prod_{\{ E_{z}\}} exp\{-\beta
(A_{0}(P_{i}) + A_{0}(P_{i+1}) +
A_{int}(P_{i},P_{i+1} ) \}  \eqno(13) $$
and the partition function as

$$  Z(\beta) = \sum_{\{ P_{i}\}~on~\{E^{i}_{z}\}}
\prod_{i} exp\{-\beta (A_{0}(P_{i}) + A_{0}(P_{i+1}) +
A_{int}(P_{i},P_{i+1} ) \} , \eqno(14) $$
where an independent summation is extended over all
closed polygons on the different planes $E_{z}$.

If we define transition amplitude from configuration $P_{i}$ on
the plane $E^{i}_{z}$ to the configuration $P_{i+1}$ on the plane
$ E^{i+1}_{z} $ as

$$ K(P_{i},P_{i+1}) =exp\{-\beta ( A_{0}(P_{i}) + A_{0}(P_{i+1}) +
A_{int}(P_{i},P_{i+1}) \}, \eqno(15)$$
then

$$ Z(\beta)=
\sum_{\{ P_{i} \}}~ \prod_{i}~K(P_{i},P_{i+1}) \eqno(16)$$
which we can interpret as the propagation of the polygon-loop
in the z-direction.

To compare with 3D Ising system
\cite{fradkin,itzykson,casher,polyakov,marinari,orland}
we will represent the energy of the interface in the form

$$S(M) = \frac{1}{2} \sum_{\{ E\}} l(E)  = \sum_{\{ E_{z} \}}
l(P_{i}) + s(P_{i}) + s(P_{i+1})
- 2 \cdot s(P_{i} \cap P_{i+1}) \eqno(17)$$
with the free action $A_{0}$ which is proportional to the
total $\it{lenght}$ plus oriented $\it{area}$

$$A_{0} = l(P)/2 + s(P) \eqno(18)$$
and interaction which is now proportional
to $\it{overlapping~area}$ of the polygon loops

$$A_{int} = -2\cdot s(P_{i} \cap P_{i+1}) \eqno(19)$$
where $s$ denotes the area. We see, that compared
with (8), the interaction is much more stronger.

Description of the spin systems in terms of interface
factors part of the symmetry: two spin configurations
connected by the global $Z^{2}$ transformation correspond
to the same surface. Description in terms
of polygon loops (8a,b,c) factors the number of surface
configurations by the factor $2^{N}$, when
$\kappa =0$. This fact is connected with the layer
symmetry of the system when $\kappa =0$ (see section 3).

The 3D Ising ferromagnet does not have this symmetry,
therefore in
(17),(18) and (19) we have to ascribe consistent orientation
to every polygon loop on the $E_{z}^{i}$ planes and then
to sum over $2^{N}$ different orientations.

{\Large{\bf5.}}Using (8a,b,c) let us define

$$A(1,2)=A_{0}(1) + A_{0}(2) + A_{int}(1,2) \eqno(20)$$
where $P_{i} \equiv i$ so that (15) becomes equal to

$$ K(1,2)=e^{-\beta~ A(1,2)}. \eqno(21)$$
Let us consider intermediate summation over all
polygon configurations between $E^{1}_{z}$ and $E^{3}_{z}$

$$\sum_{\{ 2\}} K(1,2)K(2,3) =
\sum_{\{ 2\}} e^{-\beta~ A(1,2)-\beta~ A(2,3)} \eqno(22)$$
and represent the result in the form

$$e^{-\beta~A(1,3)} \cdot F(1,3),    \eqno(22a)$$
where

$$F(1,3) =
\sum_{\{ 2\}} e^{-2\cdot \beta~A_{0}(2)}~~V(1,2,3), \eqno(23)$$
and

$$V(1,2,3) = e^{- \beta~(-A_{int}(1,3) +
A_{int}(1,2) + A_{int}(2,3))}. \eqno(24)$$
The last term in (24) represents the  interaction
which is proportional to the overlapping lenght.

{\Large{\bf 6.}}To simplify the system and to have crude
approximation   we will ignore the interaction term (24).
In that case $F(1,3)$ does not depend on the loops 1 and 3.
Therefore denoting this amplitude simply by $F$ we will have

$$ F = \sum_{ \{ P \}} e^{-2\beta \cdot
A_{0}(P)}~=~\sum_{ \{ P \}} e^{-\beta(k(P)+2l(P))}\eqno(25) $$
and that the partition function reduces to two-dimensional
partition function

$$ Z_{0}(\beta)=F^{N}. \eqno(26) $$
So in this approximation the partition function
coincides with the two-dimensional model (25)
with the energy which is proportional to the
sum of the lengths of the loops plus the
total curvature. This system coincides with
the sum of the 2D Ising model
$Z(\beta) = \sum_{\{ P\}}e^{-\beta \cdot l(P)}$
and of the 2D gonimetric walks \cite{wegner1,savvidy}
$Z(\beta) = \sum_{\{ P\}}e^{-\beta \cdot k(P)} .$
In short one can say that in this approximation the system
is equivalent to 2D Ising model in
which the paths are weighted by the total curvature.
Our aim is to evaluate (14) in this approximation.

{\Large{\bf 7.}} The two-dimensional topological model (2),
2D gonimetric walk, has the partition function (3)
and is in the disordered regime \cite{savvidy}.

For the three-dimensional case,
we obtain an additional
perimeter term to gonimetric walks (8b), (25)
and this results to the change of the phase
structure of the system in three
dimensions. We can expect that because of the perimeter
term, the linear system (6), (8) will show the
phase transition in 3D which
should be of the same nature as it is in the
2D Ising ferromagnet.

To find partition function one has to represent the corresponding
weights in terms of eight-vertex model \cite{baxter,fan,wu1,fanwu}.
The 2D Ising system has the weights

$$\omega_{1}=1,~~\omega_{2}=
w^{4},~~\omega_{3}=\omega_{4}=\omega_{5}=\omega_{6}=
\omega_{7}=\omega_{8}=w^{2},\eqno(27)$$
where $w = exp(-\beta J)$ and
$\omega_{\xi}= exp(-\beta \epsilon_{\xi})$ and
$\epsilon_{\xi}$ is the energy assigned to the
$\xi$th type of vertex configuration
($\xi_{i}=1,..,8$) $\epsilon_{1}=0,~~ \epsilon_{2}
=4J,~~ \epsilon_{3}= \epsilon_{4}= \epsilon_{5}=
\epsilon_{6}=\epsilon_{7}=\epsilon_{8}=2J.$
The weights of the gonimetric walks with $\kappa = 0$
are \cite{wegner1,savvidy}

$$\omega_{1}=\omega_{2}= \omega_{3}=\omega_{4}
=1,~\omega_{5}=\omega_{6}=
\omega_{7}=\omega_{8}=\omega ,\eqno(28)$$
where $\omega = exp(-\beta \Theta(\pi /2))$ and
$\epsilon_{1}= \epsilon_{2}=
\epsilon_{3}= \epsilon_{4}=0,~~ \epsilon_{5}=
\epsilon_{6}=\epsilon_{7}=\epsilon_{8}=\Theta(\pi /2).$
With this notations the system (25), (26) has the weights

$$\Omega_{1}=1,~~\Omega_{2}= w^{4},~~\Omega_{3}=
\Omega_{4}=w^{2},~~\Omega_{5}=\Omega_{6}=
\Omega_{7}=\Omega_{8}=\omega \cdot w^{2}.\eqno(29)$$
Following \cite {savvidy,green,hurst1,fan,fanwu}
one can rewrite the partition function (25) as a product of
the fermion operators and obtain

$$-\beta f^{0}_{gonihedric}= \frac{1}{8\pi^{2}}
\int_{0}^{2\pi} \ln [(1 + w^{4})^{2} -
4w^{8}\omega^{2}(1 - \omega^{2})
+4w^{4}(1-\omega^{2})\cos\theta \cos \phi $$
$$+2(w^{6}+w^{2}-2w^{6}\omega^{2})(\cos\theta +\cos\phi)]
d\theta d\phi , \eqno(30)$$
which confirms our expectation.
One can also compute the partition function with the
Kac and Ward \cite{kac} combinatorial approach.
To each oriented closed polygon one can correspond a term of the
matrix determinant and vice versa. For that we should define the
matrix $A((i,j)(X,Y))$, where $(i,j)$ - indicates which points
of the lattice are connected by the polygon loop and
$X = (Right,Left,Down,Up) \equiv (R,L,D,U)$ -
indicate the direction in which the bond joining the vertices
of the loop is traversed and $Y =(R,L,D,U)$ -
subsequent direction of "motion".
In our case $RR=LL= UU=DD=w,~~RU=LD=UL=DR=
= w \omega \alpha,~~RD=LU=DL=UR=w\omega \alpha^{-},$
where $\alpha = exp(-i\pi /4)$ \cite{kac}. The computation of
the determinant can be easily done and coinside with the
expression (30). We conclude from this that only boundary
terms can change this result and that the system describes
the propagation of almost free fermionic string.

{\Large{\bf8.}}We have found also the dual system to (2a),
high temperature expansion of which equally well describe
those surfaces. The high temperature expansion of (2a) is

$$ Z(\beta) = \sum_{\{ \sigma \}}
\prod_{plaquetts} ch \beta \cdot
\{ 1 + th \beta \cdot (\sigma \sigma
\sigma \sigma ) \}. \eqno(31)$$
Opening the brackets and summing over $\sigma$
one can see that only
such terms produce nonzero contribution which contain an
even number of plaquettes on every given vertex, therefore

$$Z(\beta) = (2 ch \beta )^{3N^{3}}
\sum_{ \{ \Sigma \} } (th\beta)^{s(\Sigma)}~~, \eqno(32)$$
where the summation is extended over
all surfaces $\{ \Sigma \}$
with an even number of plaquettes at any given vertex.
The $s(\Sigma)$ is the number of
plaquettes of $\Sigma$, e.g. the area of
the surface.

Let us attach plaquette variables $U_{P}$ to each plaquette
$P$ of $Z^{3}$

$$ U_{P} = -1~~~~ if~~ P
\in~~ M ~~~~and~~~~U_{P} = 1~~~~if~~\not\in M \eqno(33)$$
The constraint on the plaquette variables $U_{P}$
in every vertex

$$ \prod_{12~plaquettes~incident~to~vertex}
U_{P}=1, \eqno(34)$$
uniquely characterizes our set of surfaces $\{ \Sigma \}$.
Now one can introduce the group structure on this set
of surfaces $\{ \Sigma \}$.
Let us consider two surfaces $\Sigma^{1}$ and
$\Sigma^{2}$ and denote their plaquette variables
as $U^{1}_{P}$ and $U^{2}_{P}$ respectively.
Let us define group product of these two surfaces as

$$U_{P} = U^{1}_{P} \cdot U^{2}_{P}. \eqno(35)$$
According to this definition the set of surfaces
$\{ \Sigma \}$ (34) forms an Abelian group $G$.
The whole group $G$ is a direct product
of the local groups $G_{\xi}$. This group $G_{\xi}$
has four elements-elementary surfaces,
$G_{\xi}=\{e(\xi), g_{\chi}(\xi),
g_{\eta}(\xi), g_{\varsigma}(\xi)\}$
with the multiplication table
$e \cdot g_{\chi,\eta,\varsigma} =
g_{\chi,\eta,\varsigma};~g_{\chi} \cdot g_{\chi}
= g_{\eta} \cdot g_{\eta} =
g_{\varsigma} \cdot g_{\varsigma}
=1;~g_{\chi} \cdot g_{\eta}= g_{\varsigma},$
which follows from the fact that elementary
surfaces are match box surfaces with different
orientations and from the multiplication law (35).

Any set of elementary surfaces $e, g_{\chi},
g_{\eta}, g_{\varsigma}$ distributed independently
over the lattice $Z^{3}$ describes some
allowed surface $\Sigma$ and any given surface from
$\{ \Sigma \}$ (34) can be
uniquely decomposed into the product of $G_{\xi}$

$$\Sigma = \prod_{\xi} \cdot G_{\xi}. \eqno(36)$$
This approach allows to describe the original
surface $\Sigma$ in
terms of a new independent variable
$G_{\xi}=\{e(\xi), g_{\chi}(\xi),
g_{\eta}(\xi), g_{\varsigma}(\xi)\}$ which should be
attached to center of the cube $\xi$ of the original lattice
$Z^{3}$ or which is the same to the vertices
$\xi$ of the dual lattice
$Z^{\star~3}$.

The group $G_{\xi}$ is an Abelian group of the fourth
order and therefore has four one-dimensional
irreducible representations
$E =\{1,1,1,1\},~R^{\chi}=
\{1,1,-1,-1\},~R^{\eta}=
\{1,-1,1,-1\},~R^{\varsigma}=\{1,-1,-1,1\}$
which express algebraically the "matchbox spin" variable
$G_{\xi}$.

The dual Hamiltonian is nonhomogeneous in the directions
$\chi$,~$\eta$~and $\varsigma$

$$H_{dual} = \sum_{\xi} H_{\xi,\xi + \chi} +
H_{\xi,\xi +\eta} + H_{\xi,\xi + \varsigma}, \eqno(37)$$
where $\chi$,~$\eta$~and $\varsigma$ are unit vectors
in the corresponding directions of the dual lattice and

$$H_{\xi,\xi + \chi} \equiv H(G_{\xi},G_{\xi + \chi})
= - R^{\chi}(\xi) \cdot R^{\chi}(\xi + \chi),$$
$$H_{\xi,\xi + \eta} \equiv H(G_{\xi},G_{\xi + \eta})
= - R^{\eta}(\xi) \cdot R^{\eta}(\xi + \eta),$$
$$H_{\xi,\xi + \varsigma} \equiv H(G_{\xi},G_{\xi + \varsigma})
= - R^{\varsigma}(\xi) \cdot R^{\varsigma}(\xi + \varsigma).
\eqno(38)$$
The partition function of the dual system (37),(38)
can be written in the form $  Z(\beta^{\star}) = \sum_{\{ G_{\xi} \}}
exp\{-\beta^{\star}  H_{dual} \}$. One can check now,
that the high temperature expansion
of the dual system (37), (38) indeed coinsides with the
low temperature expansion of the original system (2a)
and provides a new realization of gonihedric system on
the lattice. This dual representation
can be compared with the dual representation of the
3D Ising ferromagnet in terms of 3D Gauge Wegner system
\cite{wegner}.

We gratefully acknowledge conversations with J.Ambjorn,
D.Gross, B.Durhuus, E.Floratos, R.Flume,
R.Schneider and F.Wegner.

One of the authors (G.K.) is thankful to H.Nielsen for
discussions and hospitality at Niels Bohr Institute; to
E.Paschos for useful comments and support during his
stay at Dortmund University; to E.Binz for
helpful conversations at Mannheim University and
to E.Marinari for the reference \cite{marinari}.

This work was sponsored in part by the Danish Natural
Science Research Council.

\vfill
\end{document}